\documentclass[12pt]{iopart}
\usepackage{epsfig}
\usepackage{graphicx}

\begin{document}

\def\alt{\raise0.3ex\hbox{$\;<$\kern-0.75em\raise-1.1ex\hbox{$\sim\;$}}}
\def\agt{\raise0.3ex\hbox{$\;>$\kern-0.75em\raise-1.1ex\hbox{$\sim\;$}}}
\def\d{{\rm d}}
\newcommand{\be}{\begin{equation}}
\newcommand{\ee}{\end{equation}}
\newcommand{\bea}{\begin{eqnarray}}
\newcommand{\eea}{\end{eqnarray}}
\newcommand{\vv}{\,\,\, ,}
\newcommand{\pp}{\,\,\, .}

\begin{flushright}
{\large \tt DSF/30/2008, IFIC/08-67}
\end{flushright}

\title{Sensitivity on Earth Core and Mantle densities using Atmospheric Neutrinos}

\author{E. Borriello$^{1,2}$, G. Mangano$^{2}$, A. Marotta$^{1,2}$, G. Miele$^{1,2}$, \\
P. Migliozzi$^{2}$, C.A. Moura$^{2,4}$, S. Pastor$^{3}$, O.
Pisanti$^{1,2}$, and \\
P. Strolin$^{1,2}$ }

\address{$^1$ \,Dipartimento di Scienze Fisiche, Universit\`a di Napoli ``Federico
II'', Complesso Universitario di Monte S.Angelo, Via Cinthia,
80126, Napoli, Italy}

\address{$^2$ INFN - Sezione di Napoli - Complesso Universitario di Monte
S.Angelo, Via Cinthia, 80126, Napoli, Italy }

\address{$^3$ Instituto de F\'{\i}sica Corpuscular (CSIC-Universitat de Val\`encia), Ed.\
Institutos de Investigaci\'on, Apartado de Correos 22085, E-46071
Val\`encia, Spain }

\address{$^4$ Instituto de F\'{\i}sica ``Gleb Wataghin'' - UNICAMP 13083-970 Campinas SP, Brazil }

\begin{abstract}
Neutrino radiography may provide an alternative tool to study the very
deep structures of the Earth. Though these measurements are unable to
resolve the fine density  layer features, nevertheless the information
which can be obtained are independent and complementary to the more
conventional seismic studies. The aim of this paper is to assess how well
the core and mantle averaged densities can be reconstructed through
atmospheric neutrino radiography. We find that about a $2\%$ sensitivity
for the mantle and $5\%$ for the core could be achieved for a ten year
data taking at an underwater km$^3$ Neutrino Telescope. This result does
not take into account systematics related to the details of the
experimental apparatus.
\end{abstract}


\maketitle

\section{Introduction}
Earth tomography with high--energy cosmic neutrinos is quite an old idea
\cite{tomo1}-\cite{GonzalezGarcia:2007gg} that seems to provide a viable
independent determination of the internal structure of our planet. In
particular, cosmic neutrinos with an energy of a few TeV have an
interaction length of the order of the Earth radius and thus sample the
density profile along the path. Detecting at a km$^3$ Neutrino Telescope
the flux of emerging charged leptons (mainly muons) versus the arrival
direction can be therefore, a promising approach for measuring at least
some of the features of the Earth density radial profile as recently
discussed in Ref. \cite{GonzalezGarcia:2007gg}. In particular, such a
measurement can provide information concerning the core/mantle boundary
which determines the geodynamo as well as the feeding mechanism of
hotspots at the surface~\cite{Loper}. The standard body-wave and free
oscillation studies, even though are much more precise tools than what it
can be reasonably obtained by neutrino radiography, are not free of
ambiguities related to the local nature of the seismometer arrays and
moreover, to the capacity of free-oscillation data to detect
one--dimensional structures only.

\begin{figure}
\begin{center}
\begin{tabular}{cc}
\\
\includegraphics[width=0.45\textwidth]{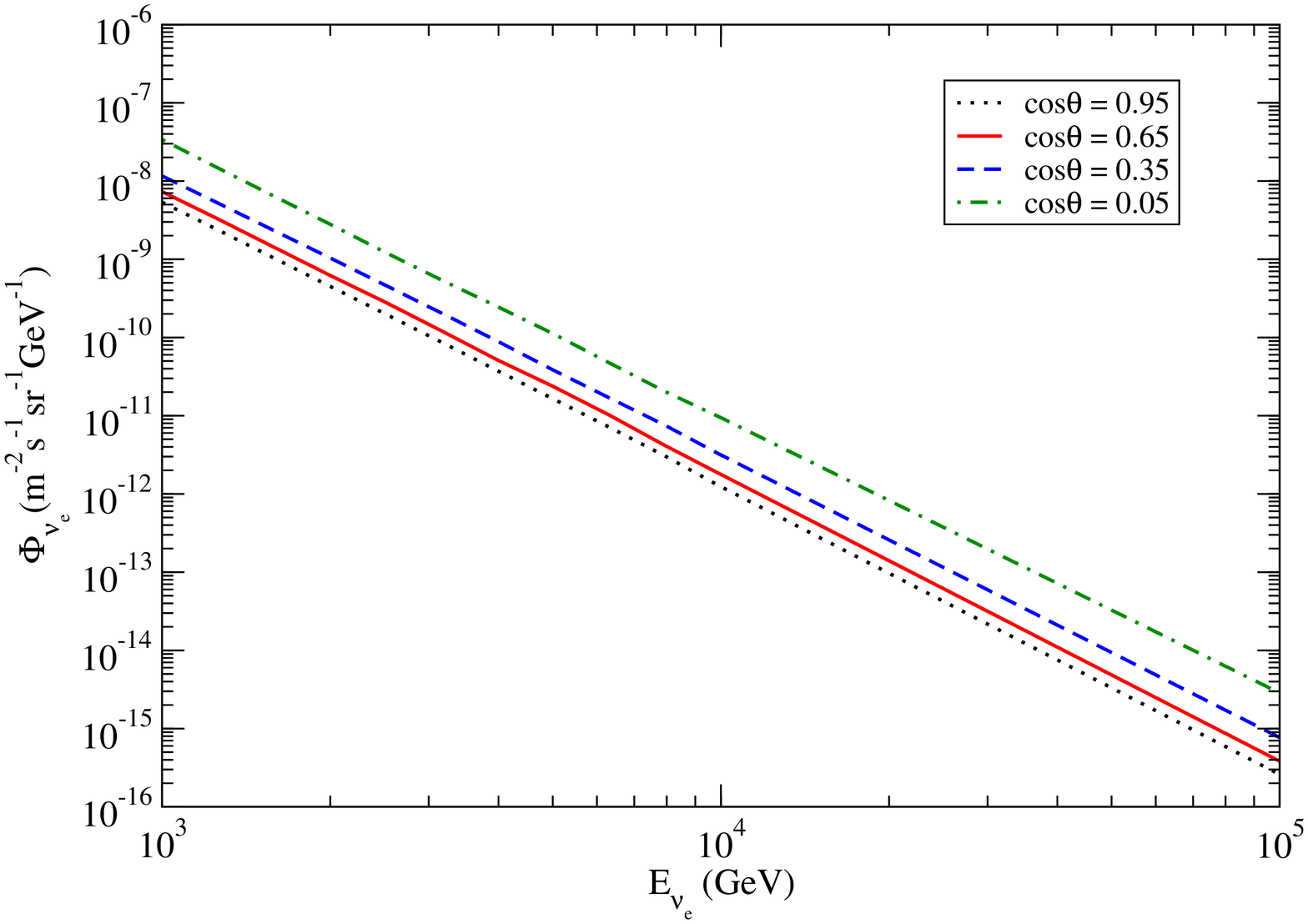} &
\includegraphics[width=0.45\textwidth]{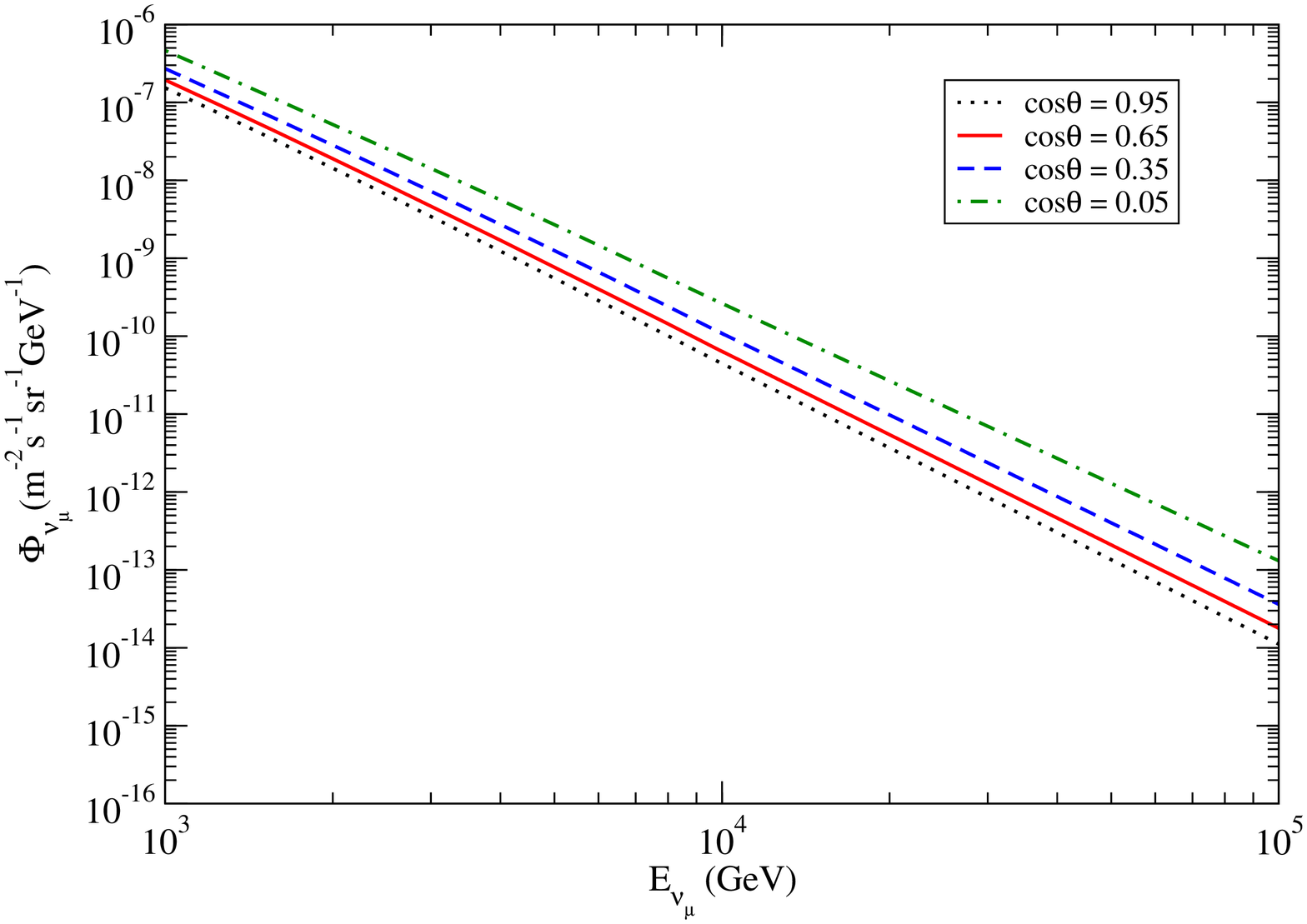} \\ \\
\includegraphics[width=0.45\textwidth]{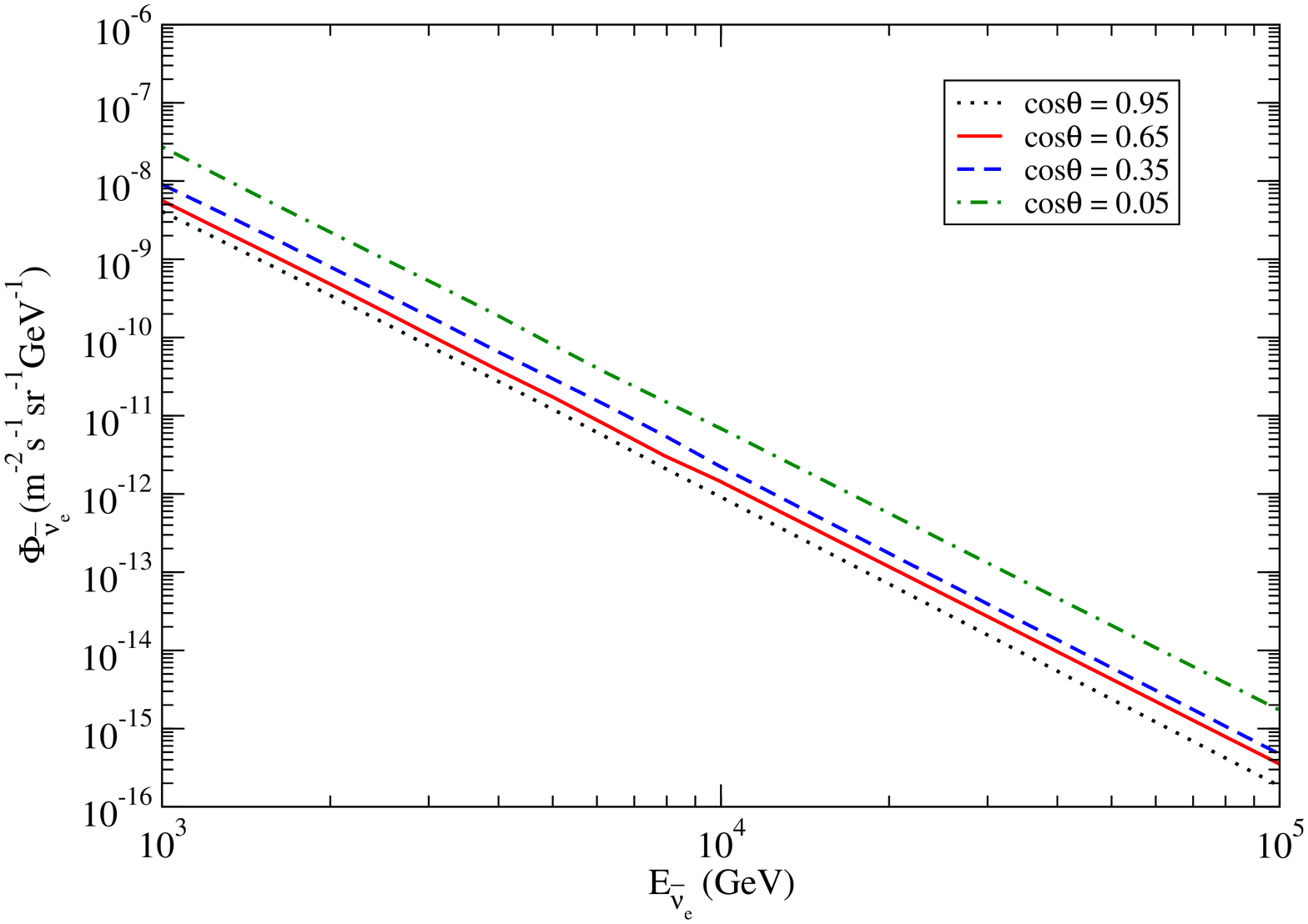} &
\includegraphics[width=0.45\textwidth]{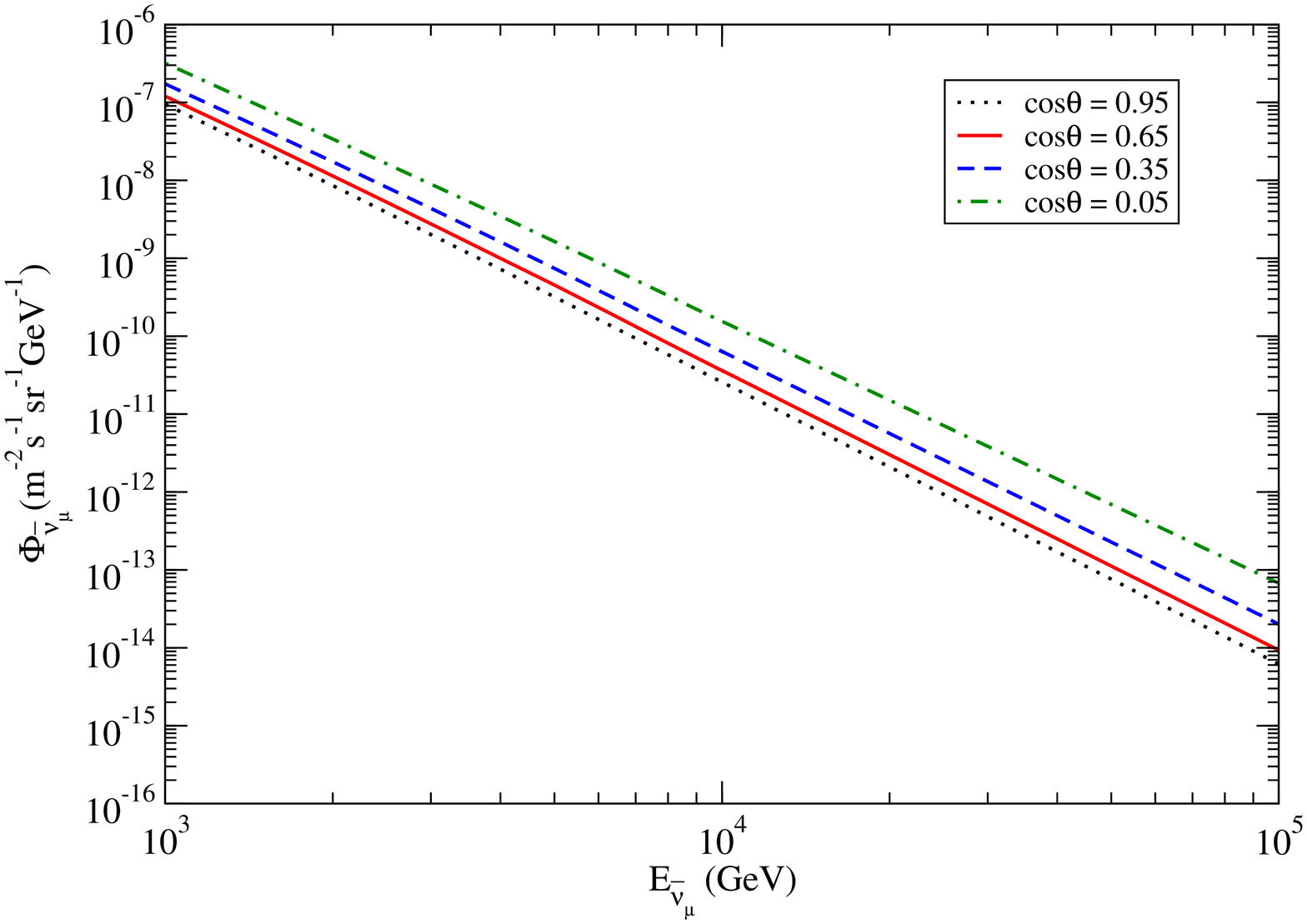}
\end{tabular}
\end{center}
\caption{Atmospheric electron and muon (anti)neutrino fluxes for different
arrival directions, with $\theta$ denoting the zenith angle. We use the
results of ~\cite{Honda:2006qj} and extrapolate the power--law behavior
above  $10^4$ GeV.}
\label{fig:nuFlux}
\end{figure}

At the energy of few TeV the neutrino flux crossing the Earth is
essentially made of Atmospheric Neutrinos (AN) which are produced in
collisions of cosmic rays with nuclei in the Earth's atmosphere. The AN
spectrum decreases very steeply with energy, as the flux is proportional
to $E^{-\gamma}$ with $\gamma \simeq 3 - 3.7$ \cite{Beacom:2004jb}. Until
energies of the order of $100$~TeV the AN are the dominant contribution to
the whole neutrino flux and this is compatible with \verb"AMANDA" bound on
extragalactic neutrinos~\cite{Beacom:2004jb,Ackermann:2004zw}. On the
other hand, at higher energies a considerable prompt neutrino flux from
the decay of heavy mesons is expected, as well as an extragalactic
neutrino  component. We are limited in our analysis by the transparency of
the Earth to the neutrinos of the lowest energies, and by their prompt and
yet uncertain extragalactic flux components at the highest energies. So,
we use the electron and muon (anti)neutrino fluxes calculated in Ref.
\cite{Honda:2006qj} by using the ``modified DPMJET-III'' in the energy
range $(10^3-10^4)$~GeV and extrapolating as a power-law behavior till
$10^5$~GeV. These fluxes show a zenith angle dependence, as shown in
Fig.~\ref{fig:nuFlux}, which should be taken into account since the
relative numbers of neutrino induced charged lepton events in the detector
in different angular bins are crucial observables for our analysis. We
notice that neutrino flavor oscillations can be neglected in the energy
range we are interested in, since they are only effective at energies
lower than $1$~TeV \cite{Martin:2003us,Beacom:2001xn}.

The detection perspectives of high energy neutrinos have received a great
interest in the past few years, in view of several proposals and R\&D
projects for Neutrino Telescopes (NT's) in the deep water of the
Mediterranean sea. In particular, the construction of the \verb"ANTARES"
NT \cite{antares} was recently completed and it is taking data.
\verb"ANTARES" \cite{antares}, as well as \verb"NEMO"
\cite{Migneco:2008zz} and \verb"NESTOR" \cite{nestor}, are involved in
research and development projects which in the future could lead to the
construction of a km$^3$ telescope as pursued by the \verb"KM3NeT" project
\cite{km3net}. Furthermore, \verb"IceCube" experimental setup, a
cubic-kilometer under-ice neutrino detector \cite{icecube} is now under
construction and already taking data. It applies and improves the
successful technique of \verb"AMANDA" \cite{amanda} to a larger volume.

\begin{figure}
\begin{center}
\includegraphics[width=0.5\columnwidth]{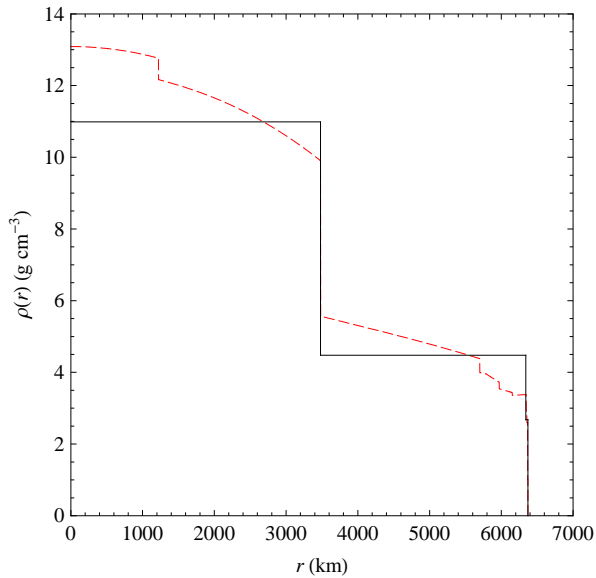}
\caption{The Preliminary Reference Earth Model radial density
~\cite{Dziewonski:1981xy} (dashed line). The solid line corresponds to a
{\it simplified} PREM (sPREM), where core and mantle densities are assumed
to be constant.}
\label{fig:prem}
\end{center}
\end{figure}

The sensitivity of a NT to the very deep geological structures is provided
by the charged lepton event rate as a function of the arrival direction,
which shows a remarkable dependence upon the adopted radial matter density
profile, as discussed in details in Ref. \cite{GonzalezGarcia:2007gg}. In
particular, in this work the authors point out the possibility to use the
arrival direction distribution of events in ten years of data taking at
\verb"IceCube", to distinguish, at $3 \sigma$ level, between the Earth
matter density profile of the Preliminary Reference Earth Model
(PREM)~\cite{Dziewonski:1981xy} and a homogeneous Earth toy model.
However, in Ref. \cite{GonzalezGarcia:2007gg} no indication of the real
sensitivity of neutrino Earth radiography to PREM parameters is reported.
In this paper we present a study of the sensitivity of an underwater NT to
Earth interior in the case of the {\it simplified} PREM (sPREM) shown in
Fig.\ \ref{fig:prem}, where core and mantle densities are assumed to be
constant and hereafter denoted by $\rho_c$ and $\rho_m$, respectively. To
assess how well these two parameters can be reconstructed, we exploit a
likelihood analysis where the reference Earth model corresponds to the
average values predicted by sPREM, $\rho_c=11.0 \, \textrm{g cm}^{-3}$ and
$\rho_m=4.48 \, \textrm{g cm}^{-3}$ with the core/mantle boundary fixed at
3450 km from the center of the Earth. We use as physical observable the
expected event rate as a function of the arrival direction, which accounts
for muon (anti)neutrino induced processes and, only for neutrino
conversions inside the NT, for the electron (anti)neutrino contribution as
well.

\section{The Monte Carlo for neutrino interaction in the Earth}

We choose as an example of undersea NT a km$^3$ detector placed at
\verb"NEMO" site, and generate a large number of tracks crossing the NT
fiducial volume (for simplicity a cube of 1 km edge placed at 100 m from
seabed) by means of a detailed Digital Elevation Map of the under-water
Earth surface, which is available from the Global Relief Data survey
(ETOPO2), a grid of altimetry measurements with a vertical resolution of 1
m averaged over cells of 2 minutes of latitude and longitude
\cite{ETOPO2}. Details can be found in
\cite{Cuoco:2006qd,Borriello:2007cs}. The simulation is performed
following tracking particles inside the rock with a maximum of 5 zones of
3 different densities, corresponding to the two regions of sPREM, as well
the crust thin layer. This leads to three possible kinds of neutrino
tracks inside the Earth: those going i) through the core, ii) through
mantle and crust, and iii) through the crust only. We inject a number of
electron and muon (anti)neutrinos for a given energy at each angular bin
according to the flux of AN given in \cite{Honda:2006qj}. We do not
consider neutrino tau contribution in this range of energy, since we are
neglecting neutrino oscillations.

The detectable events can be classified in two categories: the {\it track}
events where the charged lepton is produced outside the fiducial volume,
and the {\it contained} ones, where neutrino converts inside the NT.

\begin{figure}
\begin{center}
\includegraphics[width=0.6\columnwidth]{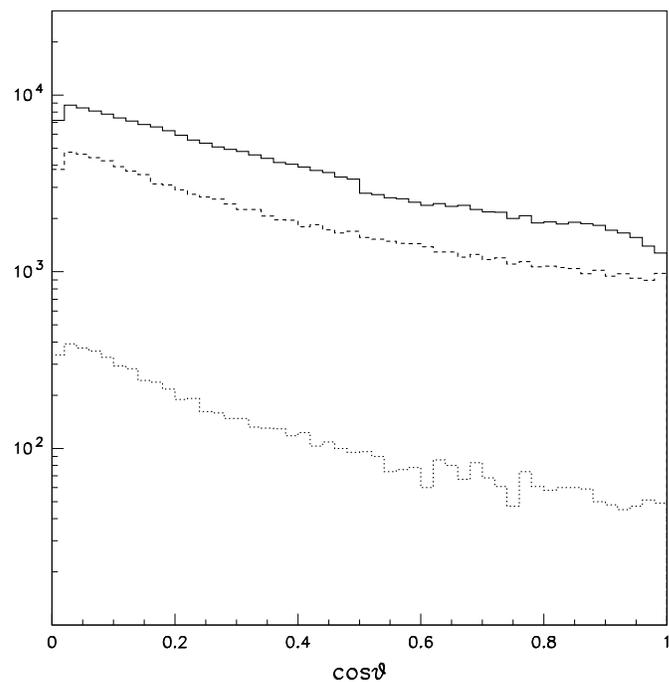}
\caption{The zenith angle distribution of neutrino induced events in a
NT for a charged lepton energy threshold of 1 TeV and the sPREM. The
up-going neutrinos  correspond to $\cos \vartheta=1$. From top to bottom,
the solid line represents track-muons, the dashed one the contained-muons
and finally the dotted line the contained-electrons.} \label{fig:angledis}
\end{center}
\end{figure}

Concerning muon neutrinos, which provide the main contribution to the
total amount of events, the Monte Carlo simulates their interaction in the
Earth and propagates the outgoing lepton. In this respect, we take into
account the phenomenon of neutrino regeneration for a neutral current
neutrino interaction. In the case of a charged current neutrino
interaction, we consider the muon energy loss in matter due to ionization,
bremsstrahlung, $e^+\,e^-$ pair production and nuclear interaction. An
energy threshold of 1 TeV is considered in counting the muons detected in
the fiducial volume and the condition of a minimal track length of 300 m
in the apparatus defines detectable events. This energy threshold value
results to be a good compromise between the need of a sufficiently large
statistics, thus a not too high lower energy threshold, and the necessity
to reduce the neutrino interaction length in order to increase the
sensitivity to Earth density profile. Contained events, which take
contributions both from electron and muon neutrinos, are treated
separately. To be conservative, in analogy to track ones we assume that
they are detected if charged lepton energy is larger than 1 TeV. This, of
course, does not take into account the amount of energy released in the
hadronic channel accompanying the charged current interaction that for
contained events could be in principle detectable. Since the contribution
of contained processes is in any case subdominant with respect to the
track one, we assume this conservative point of view which has the
advantage to make our analysis almost independent of the details of the
experimental apparatus. In Fig. \ref{fig:angledis} we show the angular
distribution of the total number of events simulated with our Monte Carlo
(at these energies the neutrino and detected charged lepton are collinear)
with $\vartheta$ the zenith angle of the emerging charged lepton, i.e.
upgoing neutrinos correspond to $\cos \vartheta=1$. Here, from top to
bottom, the solid line represents track-muons, the dashed one the
contained-muons and finally the dotted line the contained-electrons. The
electron neutrino contribution is typically smaller by one order of
magnitude than the corresponding muons due to the less abundant $\nu_e$
incoming flux (see Fig. \ref{fig:nuFlux}).

\begin{figure}
\begin{center}
\includegraphics[width=0.6\columnwidth]{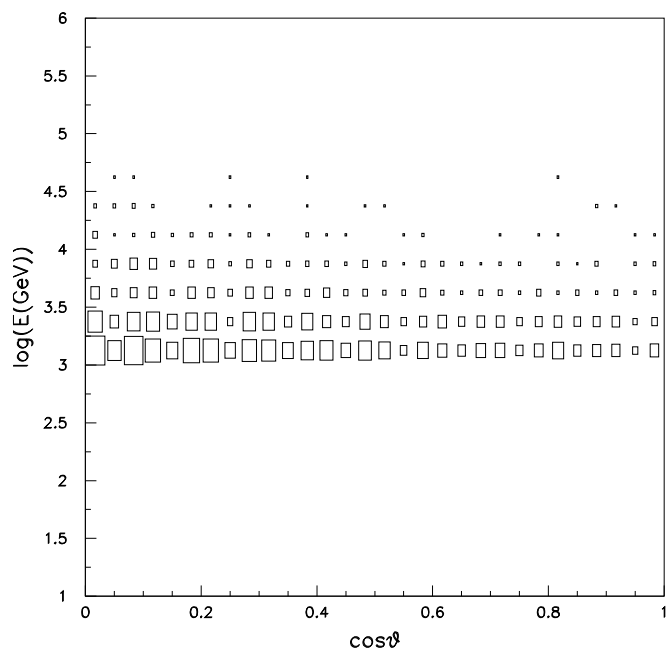}
\caption{The total amount of events as a function of both the
arrival direction and the charged lepton energy. Larger boxes correspond
to a larger number of events.} \label{fig:angleenergydis}
\end{center}
\end{figure}

In Fig. \ref{fig:angleenergydis} we show the total amount of events as a
function of both the charged lepton energy and the arrival direction. As
expected the main contribution comes from almost horizontal processes just
close to the chosen energy threshold.

\begin{table}[t]
\begin{center}
\begin{tabular}{c|c|c}
\hline\hline
$\cos\vartheta$  & sPREM  & PREM \cite{GonzalezGarcia:2007gg} \\
\hline\hline
[0, 0.17]    & 2463 & 1132 \\
\hline
[0.17, 0.33] & 1196 &  736 \\
\hline
[0.33, 0.50] &  714 &  537 \\
\hline
[0.50, 0.67] &  513 &  429 \\
\hline
[0.67, 0.83] &  306 &  359 \\
\hline
[0.83, 1.0]  &  210 &  254 \\
\hline\hline
\end{tabular}
\end{center}
\caption{The angular distribution of $\nu_\mu$ induced events in 10 years
of operation of a NT placed at NEMO site as predicted by our analysis for
the sPREM, and the PREM expectation of Ref. \cite{GonzalezGarcia:2007gg}
for IceCube. The energy threshold for muons is 10 TeV.}
\label{tab:events2}
\end{table}

In Table \ref{tab:events2} we report the angular distribution of $\nu_\mu$
induced events in 10 years of operation of a NT placed at NEMO site as
predicted by our analysis for the sPREM, and compare it with the PREM
expectation of Ref. \cite{GonzalezGarcia:2007gg} obtained for IceCube. The
energy threshold for muons has been fixed to 10 TeV since the
corresponding values for 1 TeV are absent in Ref.
\cite{GonzalezGarcia:2007gg}. As a result of this comparison the two
predictions can be considered quite in fair agreement if one takes into
account the difference in the adopted Earth density profiles and in the
level of detail of the simulation approaches. In fact, differently than in
Ref. \cite{GonzalezGarcia:2007gg} where a complete Monte Carlo of IceCube
is exploited,  we are not assuming any details of the experimental
apparatus, but simply assume a cube with a 1 km$^3$ volume and acceptance
only constrained by the requirement that charged leptons should have a
track length in the apparatus not smaller than 300 m.

\section{Results and Conclusions}

In order to carry out a sensitivity study, we vary the mantle and core
densities in a grid of $5\times4$ values: $\rho_m=\{$4.00, 4.25, 4.50,
4.75, 5.00$\}$ g cm$^{-3}$, $\rho_c=\{$9.0, 10.0, 11.0, 12.0$\}$ g
cm$^{-3}$, and calculate with the Monte Carlo the corresponding number of
events in ten years of data taking, $N_i(\rho_m,\rho_c)$, in the five
equally spaced angular bins of $\cos\vartheta$ in the interval [0,1]. We
assume an Earth radius of 6378 km and a crust with a thickness of 37 km,
while the crust density is fixed to be 2.68 g cm$^{-3}$. For each pair of
chosen values of $\rho_m$ and $\rho_c$, the radius of the core/mantle
boundary $R_c$ is then constrained by the mass of the Earth.

We then compare the counts $N_i(\rho_m,\rho_c)$ with the expected counts
$N^0_i$ for the benchmark case, $\rho_m=4.48$ g cm$^{-3}$, $\rho_c=11.0$ g
cm$^{-3}$, $R_c=3450$ km, by means of a likelihood analysis, in which the
likelihood function, ${\cal L}' (\rho_m,\rho_c,\xi,\eta) \propto
\exp(-\chi(\rho_m,\rho_c,\xi,\eta)^2/2)$, is defined using the following
expression for the $\chi^2$:
\bea \chi(\rho_m,\rho_c,\xi,\eta)^2 &=& \sum_{i=1}^5
\frac{[N_i(\rho_m,\rho_c)(1+\xi)(1-\eta\, \langle \cos\vartheta \rangle_i) -
N^0_i]^2}{N^0_i}  \nonumber\\ &+& \left( \frac{\xi}{\Delta\xi}
\right)^2 + \left( \frac{\eta}{\Delta\eta} \right)^2, \eea
where $\xi$ takes into account an overall uncertainty of the atmospheric
neutrino fluxes and neutrino interaction cross-section ($\Delta\xi=0.25$),
while $\eta$  encodes the uncertainty between horizontal and vertical
events ($\Delta\eta=0.05$) \cite{GonzalezGarcia:2007gg}.

\begin{figure}
\begin{center}
\includegraphics[width=0.6\columnwidth]{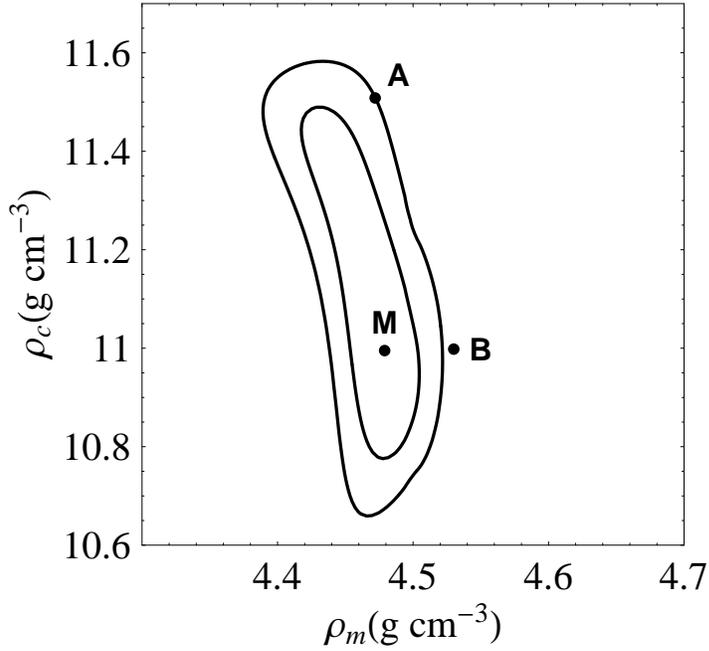}
\caption{The 68 and 95\% C.L. contours of the marginalized
likelihood function ${\cal L}$ for the measured mantle and core
Earth densities, for 10 years of data taking at a NT. The  point
denoted by M is the sPREM $\equiv (\rho_m=4.48,\, \rho_c=11.0)$ g
cm$^{-3}$, while $A\equiv (\rho_m=4.48,\, \rho_c=11.5)$ g
cm$^{-3}$ and $B \equiv (\rho_m=4.53,\, \rho_c=11.0)$ g
cm$^{-3}$ (see text and Table \ref{tab:events1}).} \label{fig:contours}
\end{center}
\end{figure}

\begin{figure}
\begin{center}
\begin{tabular}{cc}
\includegraphics[width=0.4\textwidth]{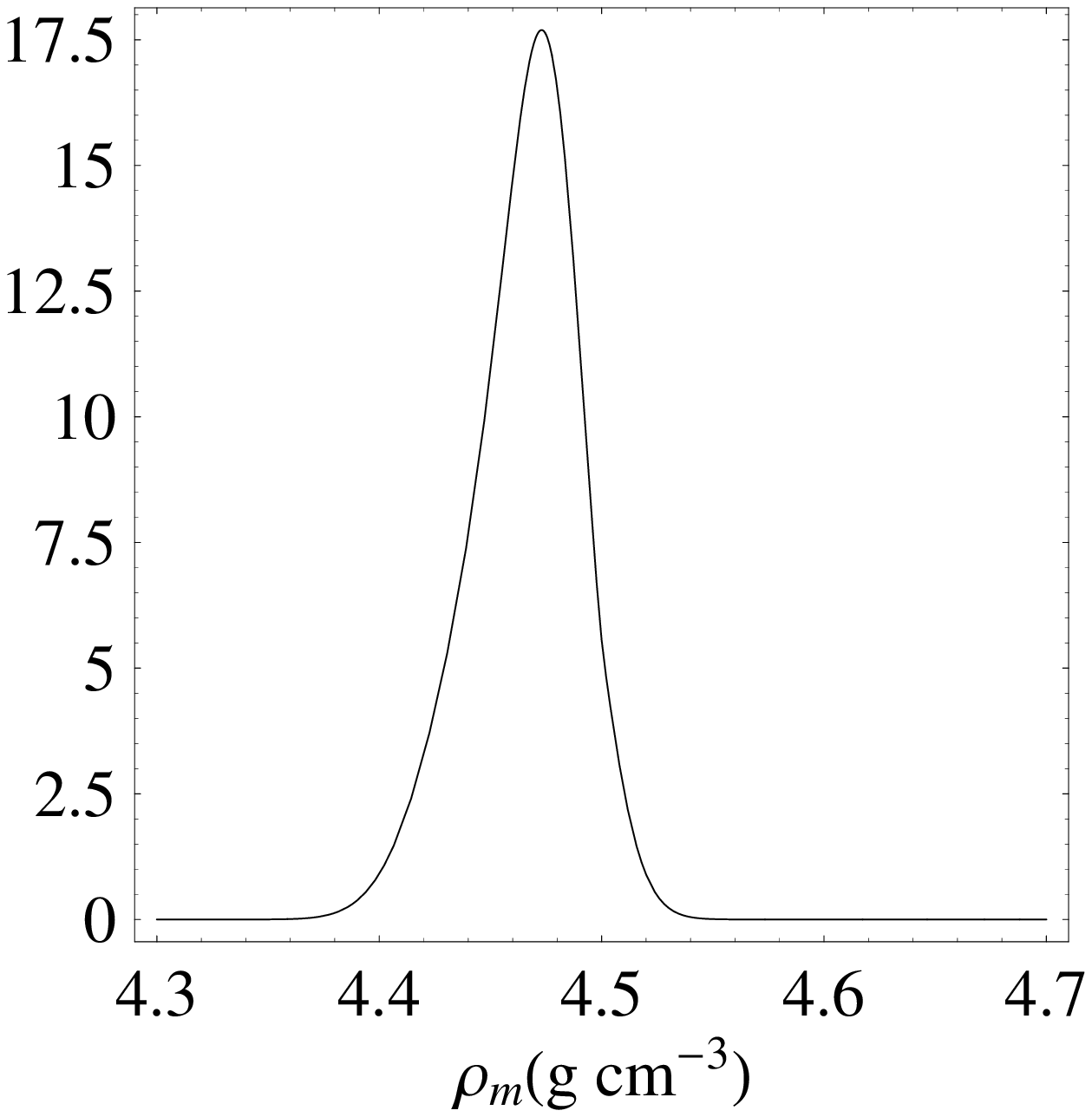} &
\includegraphics[width=0.4\textwidth]{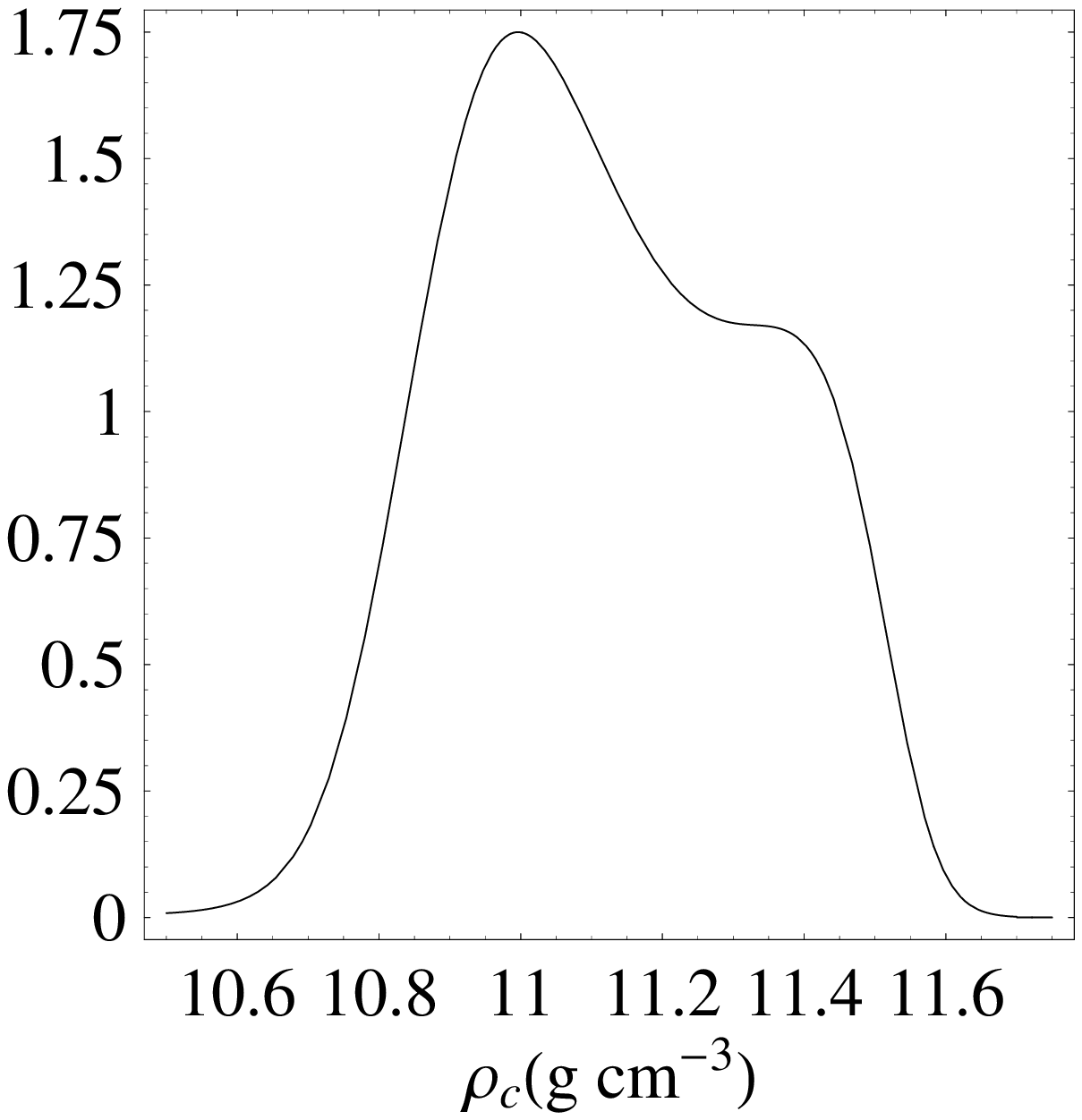}
\end{tabular}
\caption{One dimensional likelihoods for the Earth densities, $\rho_m$ and
$\rho_c$.} \label{fig:marg_rho}
\end{center}
\end{figure}

\begin{figure}
\begin{center}
\includegraphics[width=0.5\textwidth]{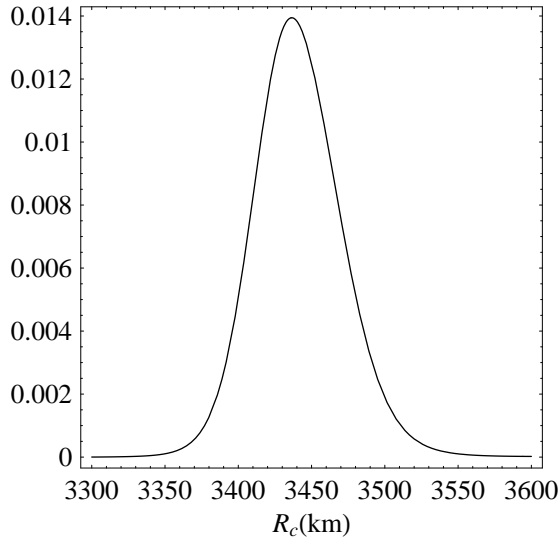}
\caption{One dimensional likelihood for the radius of core/mantle boundary,
$R_c$.} \label{fig:marg_Rc}
\end{center}
\end{figure}

We show in Fig. \ref{fig:contours} the 68 and 95\% C.L. contours of the
marginalized function with respect to $\xi$ and $\eta$,
\be {\cal L} (\rho_m,\rho_c) = \int {\cal L}'
(\rho_m,\rho_c,\xi,\eta)\, d\xi\, d\eta. \ee
By using ${\cal L} (\rho_m,\rho_c)$ we can derive the one
dimensional likelihoods reported in Figs. \ref{fig:marg_rho} and
\ref{fig:marg_Rc}
from which one obtains the ``measured" values of the densities and the
radius of core/mantle boundary at $1\sigma(2\sigma)$:
\begin{eqnarray}
\rho_m & =& 4.47\left.^{+0.02}_{-0.03}\right. \left(^{+0.04}_{-0.06}\right)~ \textrm{g cm}^{-3}\\
\rho_c & =& 11.0 \left.^{+0.3}_{-0.1}\right.
\left(^{+0.5}_{-0.2}\right)~
\textrm{g cm}^{-3}\\
R_c &=& 3440 \pm 30 \left(^{+70}_{-50}\right) ~ \textrm{km}
\end{eqnarray}

Our analysis hence suggests that a $2\%$ and $5 \%$ uncertainties (at 2
$\sigma$ level) on the averaged mantle and core densities respectively,
can be reached in a neutrino radiography campaign  with a ten years of
data taking at a typical km$^3$ NT, placed in the \verb"NEMO" site.

In order to understand how different values of $\rho_m$ and $\rho_c$ can
affect the angular distribution of events, we report in Table
\ref{tab:events1} the number of expected $\nu_\mu$ induced events in 10
years of operation of a NT placed at \verb"NEMO" site in the five bins
considered in our analysis for the sPREM, denoted by M$\equiv
(\rho_m=4.48,\, \rho_c=11.0)$ g cm$^{-3}$, and compare them with the same
expectations for two benchmark points $A$ and $B$ shown in Fig.
\ref{fig:contours} and chosen at the boundary of the 95\% C.L. region. The
energy threshold for muons is 1 TeV. This comparison illustrates the level
of sensitivity of the angular bins with respect to $\rho_m$ and $\rho_c$.
A variation of the expected number of events per bin which is typically
less than 5\% is fully compatible with a statistics larger than 10$^4$ in
ten years of running time.

\begin{table}[t]
\begin{center}
\begin{tabular}{c|c|c|c}
\hline\hline
$\cos\vartheta$  & sPREM (M)  & A & B \\
\hline\hline
[0, 0.2]   & 113436 & 113860 & 112876 \\
\hline
[0.2, 0.4] &  72393 &  75456 &  73981 \\
\hline
[0.4, 0.6] &  47334 &  48142 &  47790 \\
\hline
[0.6, 0.8] &  34105 &  34144 &  33503 \\
\hline
[0.8, 1.0] &  26781 &  27392 &  26780 \\
\hline\hline
\end{tabular}
\end{center}
\caption{Number of expected $\nu_\mu$ induced events in 10 years of
operation of a NT placed at NEMO site in the five bins considered in our
analysis for the sPREM,  denoted as M $\equiv (\rho_m=4.48,\,
\rho_c=11.0)$ g cm$^{-3}$, and for two points labeled by A$\equiv
(\rho_m=4.48,\, \rho_c=11.5)$ g cm$^{-3}$ and B$\equiv (\rho_m=4.53,\,
\rho_c=11.0)$ g cm$^{-3}$ (see Fig. \ref{fig:contours}). The energy
threshold for muons is 1 TeV.}
\label{tab:events1}
\end{table}

It is worth reminding that these results are obtained in a very simplified
PREM model, and this justifies the good level of sensitivity reachable on
$\rho_m$ and $\rho_c$ determinations. Notice also that we do not take into
account systematics related to the details of the experimental apparatus.
Sensitivity to the full PREM detailed features is much weaker, as the
number of density layers and corresponding density parameters sensibly
grow. Nevertheless, information from NT would represent an independent
confirmation of the {\it coarse grained} mantle-core transition and
provide complementary information to geophysical techniques.\\

{\bf Acknowledgments:} This work is supported by the {\it Istituto
Nazionale di Fisica Nucleare} I.S. Fa51, the PRIN 2006 ``Fisica
Astroparticellare: Neutrini ed Universo Primordiale'' of the Italian {\it
Ministero dell'Istruzione, Universit\`a e Ricerca}, the Spanish MICINN
(grants SAB2006-0171 and FPA2008-00319), by a MICINN-INFN agreement, and
by the European Union under the ILIAS project (Contract No.\
RII3-CT-2004-506222).

\end{document}